\documentclass[%
 reprint, superscriptaddress,
 amsmath,amssymb,
 aps,
]{revtex4-2}

\usepackage{graphicx}
\usepackage{dcolumn}
\usepackage{bm}

\begin{document}

\preprint{APS/123-QED}

\title{ Hot Electron Dynamics in Ultrafast Multilayer Epsilon-Near-Zero Metamaterials}

\author{Alireza R. Rashed}
\author{Bilge Can Yildiz}
\author{Surya R. Ayyagari}

\author{Humeyra Caglayan}
 \email{humeyra.caglayan@tuni.fi}
 
\affiliation{%
 Faculty of Engineering and Natural Sciences, Photonics,\\
 Tampere University, 33720 Tampere, Finland 
 }%

\date{\today}

\begin{abstract}
Realizing  remarkable  tunability  in  optical  properties without  sacrificing  speed  is critical  to  obtain  all  optical  ultrafast  devices.   In  this  work,  we  investigate the ultrafast temporal behavior of optically tunable epsilon-near-zero (ENZ) metamaterials, operating in the visible spectral range. To perform this the ultrafast dynamics of the hot electrons is acquired by femtosecond pump-probe spectroscopy and studied based on two-temperature model (2TM).  We show that pumping with femtosecond pulses changes the effective permittivity of the metamaterial more than 400 $\%$. This significant modulation is more pronounced in ENZ region and we confirm this by the 2TM. The realized ultrafast modulation in effective permittivity, along with the ultrashort relaxation time of 3.3 ps, opens a new avenue towards ultrafast photonic applications.

\end{abstract}

\maketitle



Research on optical metamaterials has seen an impressive progress during the past decade. They have created not only a better understanding of the light-matter interaction, but also a broad range of applications \cite{1Pendry, 2Jacob,3Schurig}. Among those, ENZ (epsilon-near-zero) metamaterials have become a topic of interest, because the interaction of the electromagnetic fields with such media generates particular optical properties \cite{4Engheta,5Subramania,6Silveirinha}. The unique and fascinating features of ENZ structures enable the realization of advanced optical applications such as directional light enhancement \cite{7Hajian,8Hajian}, coherent perfect absorption \cite{9Feng}, radiation pattern tailoring \cite{10Alu2007}, nonlinear fast optical switching \cite{11Wurtz}, and index of refraction sensing \cite{12Rashed}. 

The ability to tune the electromagnetic properties is crucial to develop advanced ultrafast photonic devices. The electrically tunable optical properties of transparent conducting oxide (TCO) based structures are well studied in several works \cite{13Huang,14Park,15Lee,16Vasudev,17Pradhan}. Furthermore, recent studies show that optical pumping can change drastically the optical properties of TCOs close to their ENZ wavelengths, in NIR (near-infrared) spectral range \cite{18Traviss,19Alam,20Kinsey,21Huang}. The large optically induced modification in the dielectric constant, in combination with the ultrashort response time makes TCOs a promising candidate to realize high speed tunable devices, operating near the telecommunication wavelength.  To bring the advantages and broad range of applications from NIR spectral range to visible, Papadakis \textit{et al.} have proposed a TCO based metamaterial to modify the effective optical parameters \cite{22Papadakis}. However, the performance of this device, which works by field-effect gating, is hindered due to the design complexity and low modulation speed for visible range applications. 

On the other hand, vanishing permittivity of hyperbolic metamaterials, with metal and dielectric multilayers at particular wavelengths of visible and infrared spectral ranges, make them a promising alternative to realize ENZ metamaterials \cite{23Kabashin,24Shekhar,25Sreekanth}. By changing the material composition of such metamaterials, it is possible to modify the hyperbolic dispersion and consequently the ENZ wavelength \cite{26Gao,27Engetha}. In a recent study, laser-induced tunability in the optical properties of a layered hyperbolic metamaterial is investigated in the NIR range \cite{28Kaipurath}. However, realizing a full optical tunable ENZ metamaterial with a simple design and ultrafast response for visible range applications has remained challenging.  

To control efficiently the optical ultrafast tuning of an ENZ metamaterial, it is critical to identify the temporal electron dynamics. The temporal electron dynamics and the nonlinear femtosecond response of metallic nanocrystals and metal thin films have been investigated using femtosecond pump-probe spectroscopy \cite{29Bigot1995,30Carpene,31Hamanakaa,32Bigot2000}. The temporal nonlinear response of excited systems with free electron distributions can be studied by two-temperature model (2TM) \cite{32Bigot2000,33Kaganov,34Anisimov}. The adoption of this model to ENZ systems helps to reveal the hidden aspects of the observed modulation in effective permittivity of the excited system. In a recent work, free electron dynamics of a 310 nm thick indium-tin oxide (ITO) film pumped at $\lambda_{ENZ}$ = 1240 nm, is studied using the 2TM \cite{19Alam}. The nonlinear behavior in the laser-induced structure is attributed to the thermalization of the conduction-band electrons. However, the temporal nonlinear dynamics related to the relaxation of hot electrons in an optically excited multilayer ENZ metamaterial has remained unrevealed so far, which is one of the focuses of this study.

Here, we present a tunable ENZ metamaterial based on a simple design of multilayer metal-dielectric nanostructure. The proper composition of gold (Au) and Titanium dioxide (TiO$_{2}$) thin films provides an ENZ response in the visible spectral range. We investigate the modulation of the effective permittivity of the laser-induced ENZ metamaterial based on the hot electron generation effect.  We quantitatively analyze for the first time the ultrafast temporal behavior of electron and lattice temperatures of multilayer ENZ metamaterials by 2TM. We show that the ultrafast relaxation of the excited metamaterial is determined by the electron-phonon interactions. The observed maximum change in the permittivity is confirmed by the 2TM. Our investigations reveals the potential of the metamaterials for ultrafast tunable optical devices in visible range. 



A multilayer metamaterial structure composed of 16 nm of Au and 32 nm of TiO$_{2}$ is fabricated on a  fused silica (SiO$_{2}$) substrate (Fig. 1(a)). Four pairs of metal-dielectric stack is alternatively deposited using electron-beam evaporator. The SEM image presented in Fig. 1(b) shows the cross-section view of the fabricated sample.

\begin{figure}[h]
 \includegraphics[width=\linewidth]{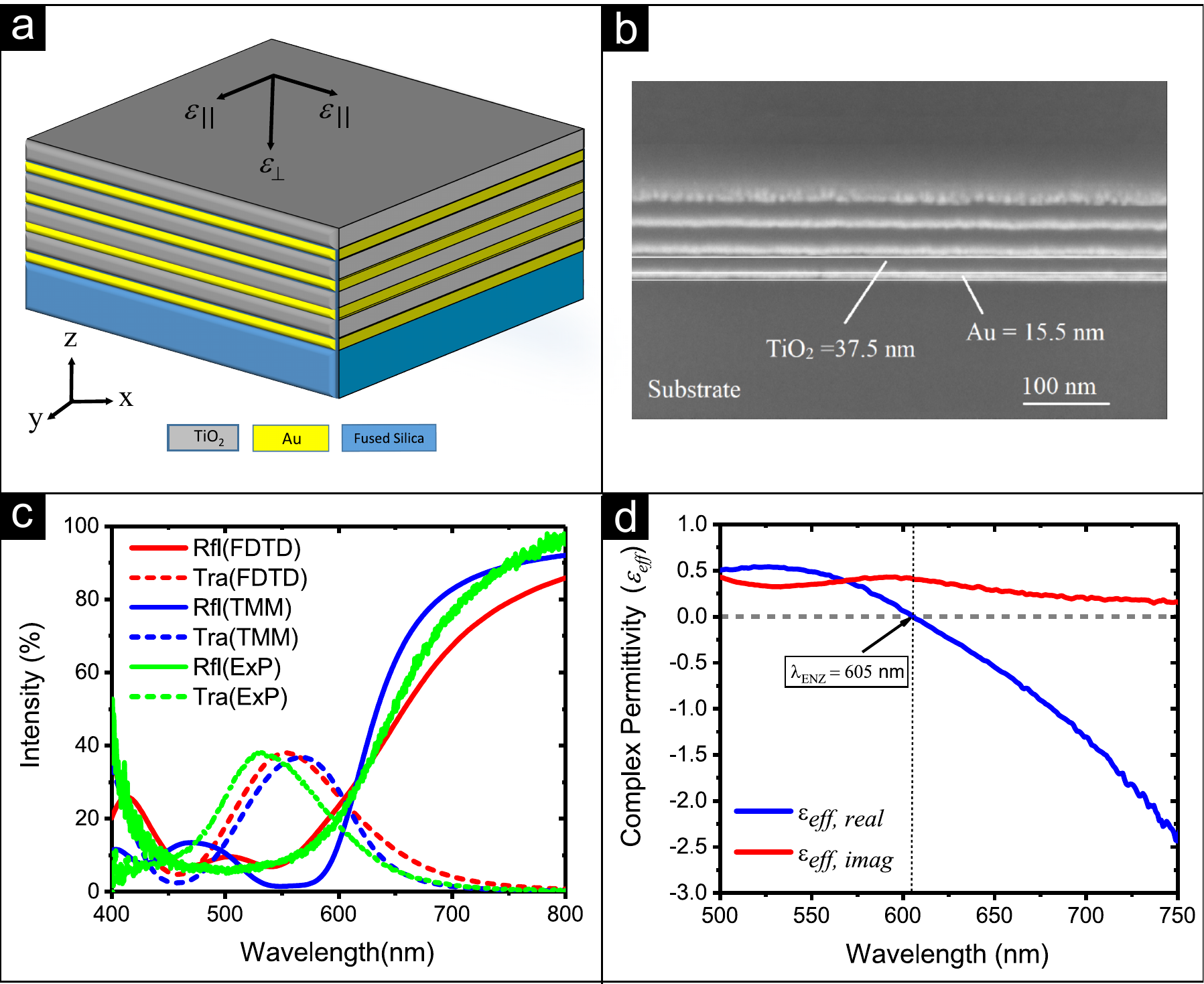}
 \caption{\label{fig:epsart} (a) Scheme for the designed multilayer nanostructure (b) The cross-sectional SEM image of the metamaterial. (c) Reflectance (solid lines) and transmittance (dashed lines) of the metamaterial; experimental (green), TMM calculations (blue) and FDTD simulations (red). (d) The calculated effective real and imaginary parts of the parallel complex permittivity.}
\end{figure}

A confocal microscope in combination with a CCD spectrometer, operating in visible spectral range, is used for acquiring the linear reflectance (R$_{lin}$) and transmittance (T$_{lin}$) of the fabricated sample.  In addition to the measurements, these parameters are calculated by using transfer matrix method (TMM), and simulated by Lumerical FDTD Solutions. In these calculations, Drude-Lorentz model is used for Au layer refractive index and the dispersion values for TiO$_{2}$ is adopted from reference \cite{35DeVore}. The experimental and simulation results show an excellent agreement (Fig. 1(c)). Figure 1(d) depicts the calculated effective parallel permittivity ($\varepsilon_{eff}$) for the fabricated metamaterial, retrieved from the measured R$_{lin}$ and T$_{lin}$ data by applying inverse TMM. We observe that the real part of the permittivity approaches to zero at 605 nm ($\lambda_{ENZ}$ = 605 nm).


In order to investigate the optical tunability of the designed metamaterial, we have measured the reflectivity, $\Delta$R/R$_{lin}$, and transmittivity, $\Delta$T/T$_{lin}$, by pump-probe spectroscopy. Here, $\Delta$R = R$_{pump}$ - R$_{lin}$ and $\Delta$T = T$_{pump}$ - T$_{lin}$, where R$_{lin}$ and T$_{lin}$ are the reflectance and transmittance in the presence of weak probe pulses (i.e., linear response), while T$_{pump}$ and T$_{pump}$ are the reflectance and transmittance in the presence of intense pump pulses. The pump signal is fixed at a single wavelength with a pulse duration of 120 fs, while the broadband probe beam (500 nm to 750 nm) is varied temporally with respect to the pump beam by using a precise delay line. We have gradually increased the pump signal wavelength from 420 nm to 500 nm with the step size of 10 nm at a constant power in order to identify the maximum modification in the transmitted ($\Delta$T/T$_{lin}$) and reflected ($\Delta$R/R$_{lin}$) probe signals. The maximum change in optical constants is observed at $\lambda_{pump}$=470 nm. As seen in Fig. 1(c), this wavelength corresponds to the maximum absorbance of the multilayer nanostructure, in which both transmittance and reflectance reach to their minimum values below the ENZ wavelength. 
\begin{figure}[h]
 \includegraphics[width=\linewidth]{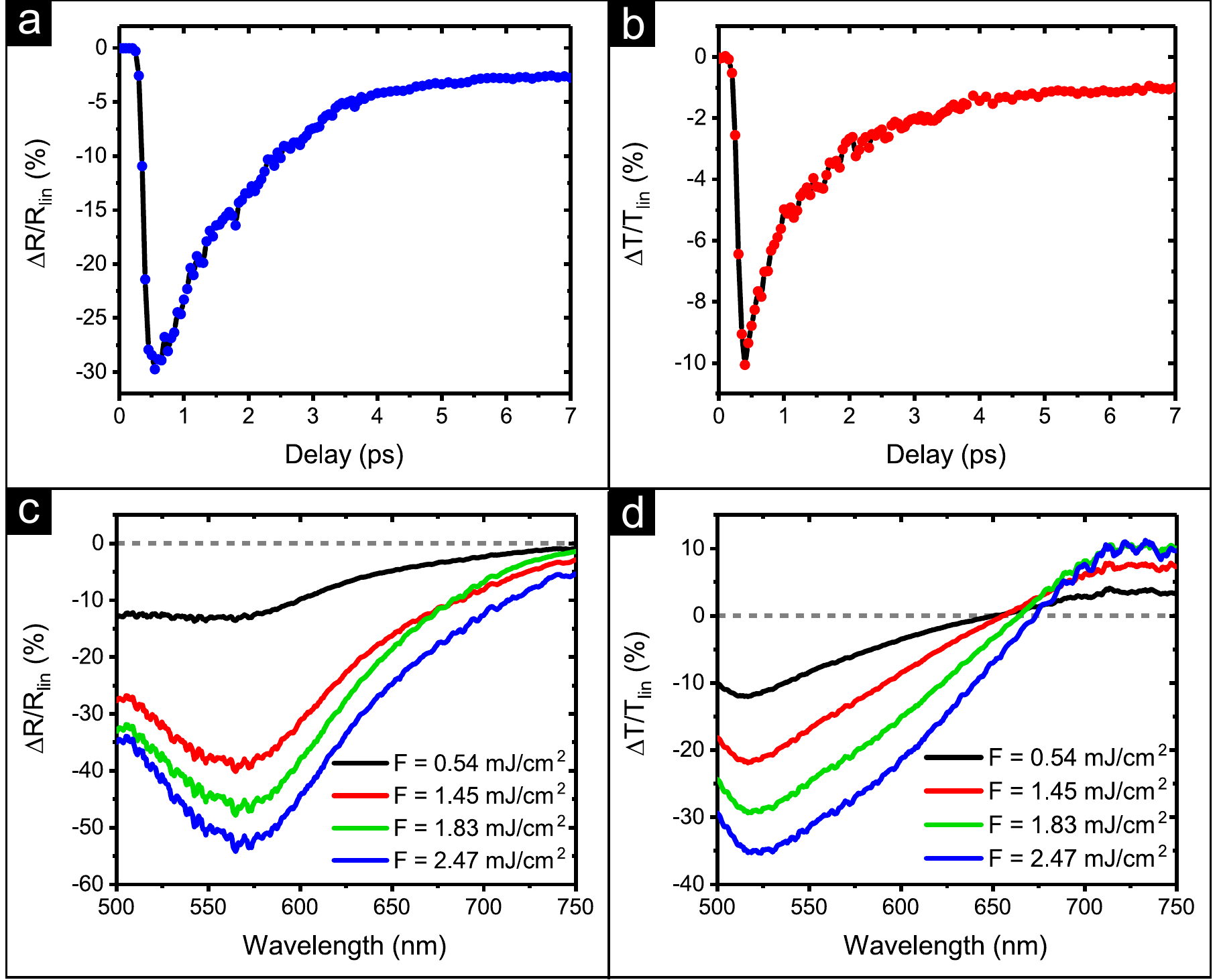}
\caption{\label{fig:epsart} (a) Reflectivity ($\Delta$R/R$_{lin}$) and (b) transmittivity ($\Delta$T/T$_{lin}$) of the pumped metamaterial as a function of time delay for the fixed pump fluence of 1.45 mJ/cm$^{2}$ at $\lambda_{pump}$=470 nm. (c) The spectra of reflectivity and (d) transmittivity as function of probe wavelength at four different fluences.}
\end{figure}

Furthermore, the probe pulses are delayed with respect to the pump pulses to investigate the ultrafast temporal dynamics of the designed system. Figure 2(a) and (b) show the change in the reflectance and transmittance of the metamaterial, pumped at fluence of F=1.45 mJ/cm$^{2}$, with respect to the time delay. The modification in the optical response reaches to a maximum few hundred femtoseconds after the pump pulse. Figure 2(c) and (d) show the spectra of reflectivity and transmittivity as a function of probe wavelength at four different fluence values of 0.54 mJ/cm$^{2}$, 1.45 mJ/cm$^{2}$, 1.83 mJ/cm$^{2}$ and 2.47 mJ/cm$^{2}$. The intensity of the probe beam changes gradually, as the pump fluence increases. At sufficient intensities of the pump beam, a significant modification in both reflectivity and transmittivity is observed. 
In Fig. 2(c) (2(d)), the negative sign of reflectivity (transmittivity) corresponds to transient absorption, whereas a positive sign corresponds to transient bleach. At the wavelengths below $\sim$ 660 nm, as a consequence of simultaneous optically induced reduction in reflectivity and transmittivity, the absorbance of the metamaterial increases. However, for the spectral range beyond 660 nm, $\Delta$T/T$_{lin}$ goes above zero, while $\Delta$R/R$_{lin}$ remains at negative values. Considering that for the bleach region of the transmittivity, the linear transmittance is almost zero, further reduction of this parameter is not allowed. Consequently, the negative values of reflectivity results  reduction in the absorbance.


The pump-induced complex $\varepsilon_{eff}$ of the designed metamaterial can be extracted by applying inverse TMM over the experimentally acquired reflectivity and transmittivity, and the R$_{lin}$ and T$_{lin}$. Although the modification of the complex permittivity is observed for a broad spectral range of the probe beam,  as shown later, this modification is much more pronounced in the ENZ region, as the occurred nonlinear effects are stronger in that region \cite{19Alam}. Figure 3(a) shows the modification of the ENZ wavelength as a function of the incident pump fluence. We observe that the pump pulse can induce 10 nm of red-shift in the ENZ wavelength, with respect to the linear case (605 nm). The reduction in the pump-induced plasma frequency around ENZ wavelength is resulted from the increase in the free-electrons of the metamaterial. In Fig. 3(b), we extract the real and imaginary parts of the pump-induced permittivity at 610 nm, as the transition from dielectric to metallic regime is pronounced around this wavelength. The total increase of $\Delta\varepsilon_{eff,real} \approx$ 0.16 ((420
\% change) is evident in Fig. 3(b), compared to linear regime. 

\begin{figure}[h]
 \includegraphics[width=\linewidth]{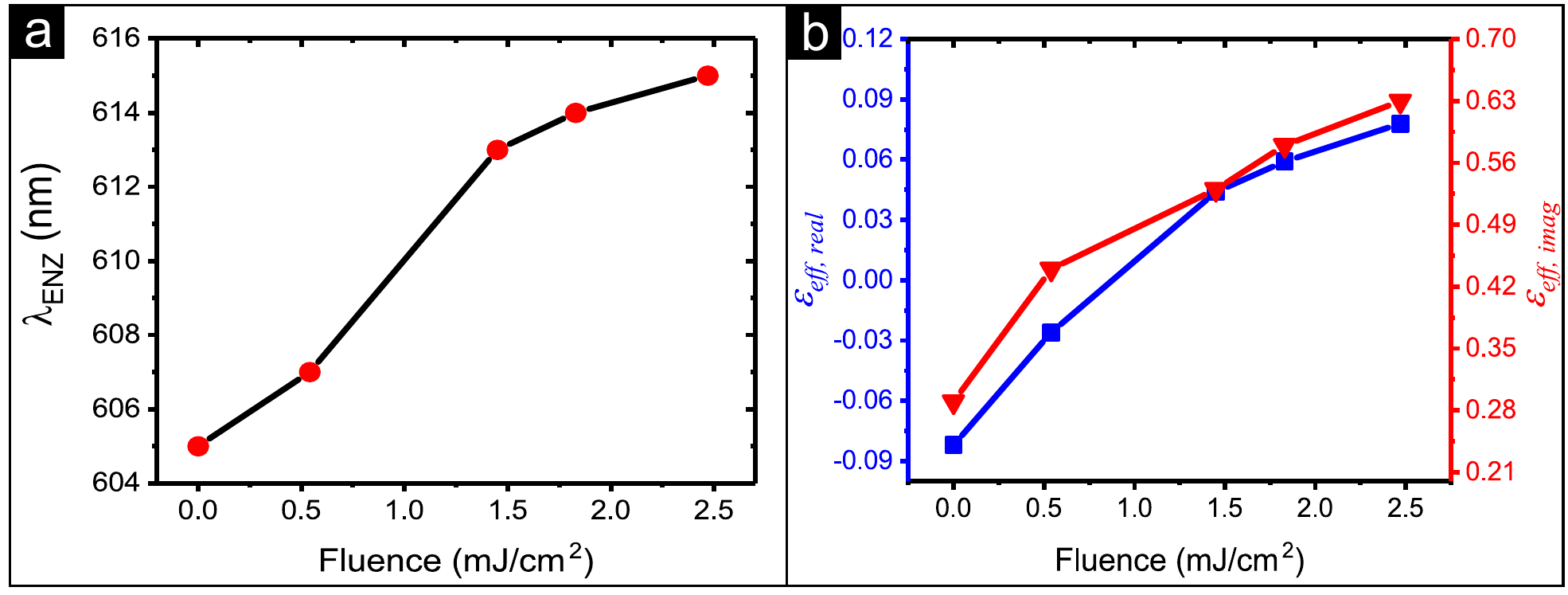}
\caption{\label{fig:epsart} (a) The tuning of the ENZ wavelength as a function of pump power. (b) Modulation of the real and imaginary parts of the pump-induced permittivity around ENZ wavelength. $\lambda_{pump}$ and $\lambda_{probe}$ are considered as 470 nm and 610 nm, respectively.}
\end{figure}


The pump pulse creates energetic non-equilibrium electrons in the conduction band, which then thermalize in few hundred femtoseconds adopting a new Fermi-Dirac distribution with a well-defined temperature \cite{36Groeneveld,37Guo}. Pump-induced changes in the permittivity are attributed to these modifications in the energy distribution, also known as hot electron generation \cite{38Caspani,39Reshef}. The ultrafast relaxation of the thermalized electrons occurs at a rate determined dominantly by the electron-phonon coupling over a timescale of a few picoseconds. To understand the observed ultrafast temporal behavior of the permittivity modulation, we model the ENZ metamaterial using a phenomenological 2TM \cite{40Chen}. We define distinct temperatures for the ionic and electronic subsystems, i.e., lattice temperature (T$_{l}$), and electron temperature (T$_{e}$), which are temporally coupled to each other by a coupling constant, G, and driven by the external pump pulse, P(t). We note that the other relaxation channels such as thermal conductivity and heat diffusion are neglected, as the timescale of interest here is 0.5-5 ps \cite{32Bigot2000}. The two coupled equations of the 2TM are the following, 
\begin{equation}
C_e(T_e)\frac{dT_e}{dt}=-G(T_e-T_l)+P(t),\
\end{equation}
\begin{equation}
C_l(T_l)\frac{dT_l}{dt}=G(T_e-T_l).
\end{equation}
Here, C$_{e}$ and C$_{l}$ denotes the specific heat capacity of the electrons and lattice, respectively and P(t) is the power of the pump pulse, absorbed by the metamaterial, which is expressed as a Gaussian distribution as following equation,
\begin{equation}
P(t)=\frac{AF}{L\tau_p}e^{-2(\frac{t-t_0}{\tau_p})^2}\
\end{equation}
where A is the absorption, F is the fluence, L is the penetration depth, $\tau_{p}$ is the pump pulse duration, and t$_{0}$  is the response delay. We use the electron specific heat capacity of Au, C$_{e}$(T$_{e}$)= 62.9T$_{e}$ J/(mK$^{2}$) \cite{41Lin}, since by pumping at 470 nm, the generation of hot electrons occur only due to the existence of the Au layers in the composition of the designed multilayer metamaterial.The lattice specific heat capacity of Au is 2.51$\times$10$^{6}$J/(m$^{3}$K), whereas that of TiO$_{2}$ is 2.77$\times$10$^{6}$ J/(m$^{3}$K) \cite{42Mehmetoglu}, which are both independent of the lattice temperature up to very high temperatures (${>}$5000 K). We take the average of the lattice specific heat capacities of the two materials with respect to their composition in the metamaterial and employ an effective specific heat capacity, C$_{l}$ = 2.68$\times$ 10$^{6}$ J/(m$^{3}$K). Source pulse duration is $\tau_{p}$ = 120 fs, and the response delay is t$_{0}$ = 0.5 fs. Absorption, A, is determined from the linear measurements, varying between 50\% to 62\% in the range of interest, 540 nm – 640 nm, in which the ENZ region lies. 

To estimate the electron-phonon coupling constant, G, the electric permittivity changes ($\Delta\varepsilon_{eff,real}$) are extracted at four different used pump fluence values (Fig. 4(a)). This is done by applying inverse TMM method on the reflectivity and transmittivity data of the excited ENZ metamaterial at the probe wavelength of 590 nm. As shown in previous studies the maximum change in reflectivity and transmittivity correspond to the highest electron temperature change, before delivering this energy to the cold lattice \cite{31Hamanakaa,43Hohlfeld,44Daraszewicz}. 
The coupling constant is determined by an iterative two-step analysis method. We start with an initial guess value for the G, and solve the 2TM equations, and obtain the electron and lattice temperatures with respect to time, for each fluence. It is possible to separate the influences of temperature changes in the lattice and electrons to the total change in the permittivity. Therefore one can parametrize experimental $\Delta\varepsilon_{eff,real}$ with the electron and lattice temperatures, i.e., $\Delta\varepsilon$(T$_{e}$, T$_{l}$)= $\Delta\varepsilon$(T$_{e}$) + $\Delta\varepsilon$(T$_{l}$). As shown in Fig. 4(a), the permittivity of the metamaterial experiences a rapid increase, reaching its maximum almost simultaneously with the pump. We correlate this maximum $\Delta\varepsilon_{eff,real}$ with the theoretical T$_{e}$ maximum, assuming T$_{l}$ is unperturbed at this particular time delay. This is a legitimate assumption, as the change in the lattice temperature is at least two orders of magnitude less than the change in the electron temperature. Similarly, one can correlate the steady-state temperature of the lattice with the experimental $\Delta\varepsilon_{eff,real}$ when the fast transient component has decayed away (i.e., after 6-8 ps). More specifically, we solve the 2TM equations for a guessed G to find maxima of T$_{e}$ and steady-state values of T$_{l}$, and match them with the experimental maxima and steady-state values of $\Delta\varepsilon_{eff,real}$ at each fluence. This matching allows us to obtain the parametrized effective permittivity change as a function of T$_{e}$ and T$_{l}$, as follows, 
\begin{equation}
\Delta\epsilon(T_e,T_l)=A_0+A_1T_e+A_2T_e^2+B_0+B_1T_l
\end{equation}
where the coefficients, A$_{i}$ and B$_{i}$ are determined from the fit functions of the  correlated $\Delta\varepsilon_{eff,real}$ and temperatures. We are now able to compare the experimental and parameterized $\Delta\varepsilon_{eff,real}$ with different guessed G values. We start with the electron-phonon coupling constant of Au, G = 3$\times$10$^{16}$ W/(m$^{3}$K) \cite{41Lin} and change it step-wisely to estimate the one for the studied metamaterial. Figure 4(b) shows the comparison of the experimental $\Delta\varepsilon_{eff,real}$ (blue line) measured at $\lambda$ = 590 nm, for the fluence, F = 2.47 mJ/cm$^{2}$, with the parametrized $\Delta\varepsilon_{eff,real}$ for different guessed G values. 

\begin{figure}[h]
 \includegraphics[width=\linewidth]{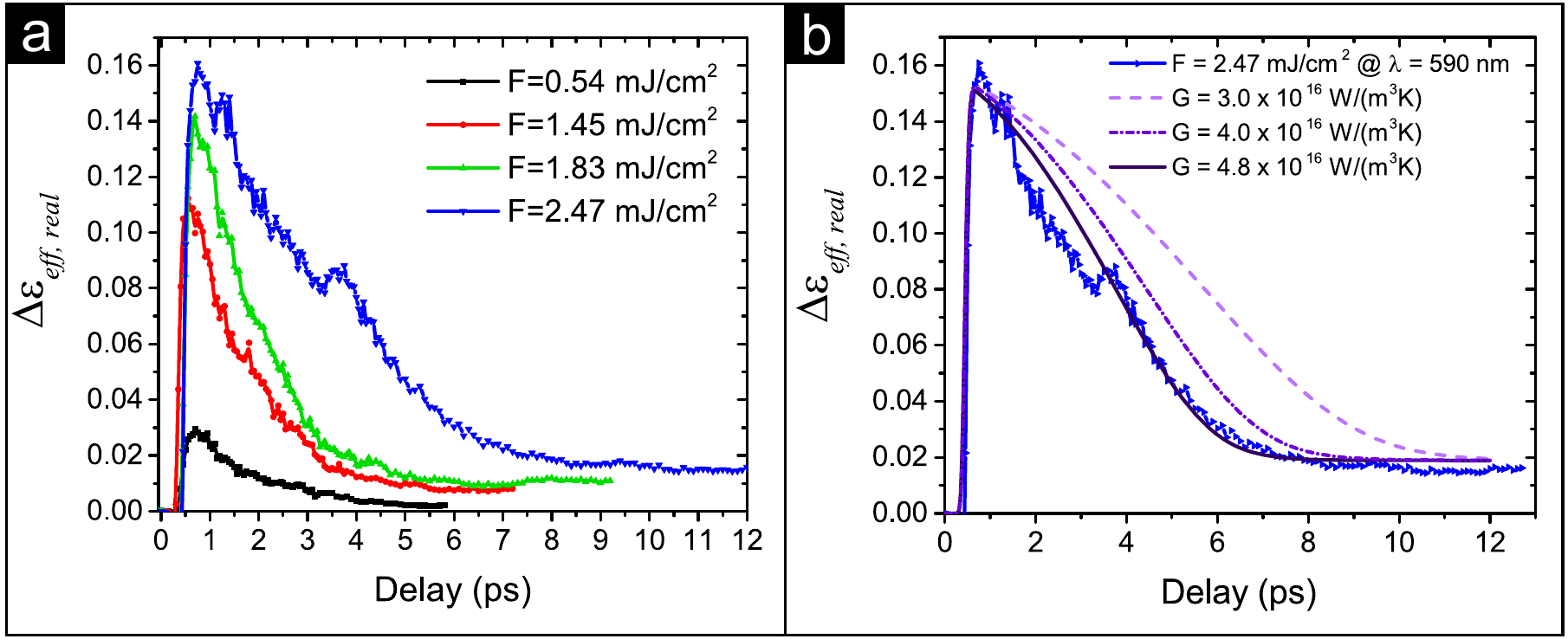}
\caption{\label{fig:epsart} (a) The electric permittivity changes ($\Delta\varepsilon_{eff,real}$) as a function of delay time extracted from the reflectivity and transmittivity data of the excited ENZ metamaterial for the probe wavelength of 590 nm. (b) The comparison of the experimental $\Delta\varepsilon_{eff,real}$ (blue), for the pump fluence of F = 2.47 mJ/cm$^{2}$, with the parameterized $\Delta\varepsilon_{eff,real}$ for three different guessed values of G parameter.}
\end{figure}

The solutions of 2TM can be used to obtain the time dependence of T$_{l}$ and T$_{e}$. The electron dynamics is governed mainly by electron–electron, electron-surface scatterings and electron-phonon interactions. Figure 5(a) shows the calculated exchange of temperature between the thermalized electrons and the lattice based on the estimated electron-phonon coupling constant. The process lasts for few picoseconds, until the hot electrons and cold lattice phonons reach to an equilibrium. Estimating the electron-phonon coupling constants for the metamaterial next, we apply 2TM in a specific spectral range. Figure 5(b) shows the maximum electron temperatures reached at different wavelengths for F = 1.83 mJ/cm$^{2}$. The spectral behaviour of the maximum electron temperature displays a peak in the ENZ region, due to the much more pronounced nonlinearity of the pumped metamaterial in this region. Thermalization and cooling down of the hot electrons are the two main processes which determine the nonlinear transient response of an excited material with femtosecond pulses. The temporal response of the system provides the information regarding the rise time and the relaxation time of the excited system. Figure 5(c) shows the analyzed ultrafast temporal response of the multilayer ENZ metamaterial to femtosecond pulses for the fluence value of F = 1.83 mJ/cm$^{2}$, based on the modification of the effective permittivity at 610 nm. The acquired 250 fs rise time in combination with 3.3 ps relaxation time are evidences for the ultrafast performance of our tunable ENZ metamaterial. 

\begin{figure}[h]
 \includegraphics[width=\linewidth]{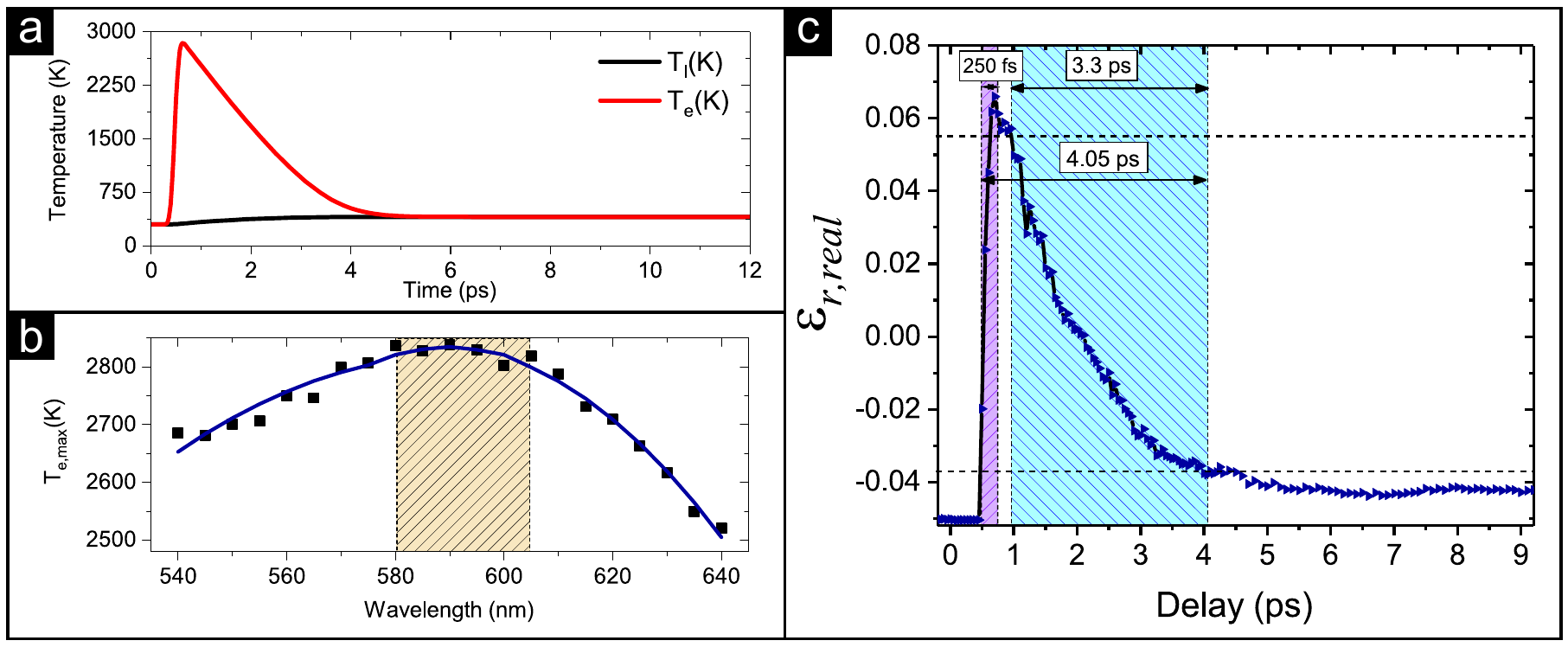}
\caption{\label{fig:epsart} (a) Calculated electron dynamics by 2TM. (b) The calculated maximum electron temperatures at different wavelengths for the pump fluence of F = 1.83 mJ/cm$^{2}$. (c) The ultrafast modification in the effective permittivity of the ENZ metamaterial. Two horizontal dashed lines show 10\% and 90\% of the maximum modification of the permittivity for calculating the rise time and relaxation (fall) time of the excited system.}
\end{figure}
In this work, we reported ultrafast optical tuning of the metal-dielectric multilayer by exploiting the ENZ regime in the visible range.  At proper energies of the optical pump pulses, we demonstrated that the effective permittivity of the ENZ metamaterial is altered considerably. The hot electron generation modifies the reflectivity and transmittivity, and hence the permittivity of the excited system. The relaxation dynamics of the hot electrons is analyzed by applying two-temperature model, based on the results of ultrafast transient absorption spectroscopy. We showed that the excited electrons' temperature reaches to maximum, resulting in the highest change of the permittivity in the ENZ region. Such pump-induced permittivity modification leads to a red-shift of the ENZ wavelength. The observed extremely-short rise and relaxation times allows the ultrafast functionality for the tunable system with the modulation speed of at least 0.25 THz. The reported results prove the possibility to realize full optical ultrafast signal processing in the visible spectral range this can pave the way for the development of practical and cost-effective optical devices.

\begin{acknowledgments}
We acknowledge the financial support of the European Research Council (Starting Grant project aQUARiUM; Agreement No. 802986), Academy of Finland Flagship Programme, (PREIN), (320165).
We thank Prof. Nikolai Tkachenko for his help to perform ultrafast pump-probe measurements.

\end{acknowledgments}

\nocite{*}

\bibliography{apssamp}

\end{document}